\documentclass{TTP_DSL2006}
\usepackage[english]{babel} 
\usepackage{array}
\usepackage{mathtext}
\usepackage{textcomp}
\usepackage{cite}
\usepackage{graphicx}
\usepackage{sidecap}
\sidecaptionvpos{figure}{c}

\newcommand{\be}{\begin{equation}}
\newcommand{\ee}{\end{equation}}
\newcommand{\ve}{\varepsilon}

\usepackage{graphicx}
\usepackage[intlimits]{amsmath}
\usepackage{amssymb}
\usepackage{exscale}
\usepackage{times}

\renewcommand{\large}{\fontsize{14}{18pt}\selectfont}
\renewcommand{\small}{\fontsize{11}{13.6pt}\selectfont}

\newcommand{\titleformat}{\sffamily\bfseries \large}				
\newcommand{\authorformat}{\sffamily \large}						
\newcommand{\keywordsformat}{\noindent \small \sffamily}			
\newcommand{\abstractformat}{\noindent \textbf}						
\newcommand{\contentformat}{\rmfamily \normalsize\vspace{18pt}}		
\newcommand{\email}{\sffamily \small \vspace{-8pt}}					
\renewcommand{\subsection}{\textbf}									

\begin{document}

\selectlanguage{english} 

\title{\titleformat Magnetic states of heterophase particle in the field of mechanical stresses}

\author{\authorformat Leonid Afremov \inst{1} and Yury Kirienko \inst{2}$^{,*}$}

\institute{\sffamily Far-Eastern Federal University, Vladivostok, Russia}
\maketitle

\begin{center}
\email{$^{1\,}$afremovl@mail.dvgu.ru, $^{2\,}$yury.kirienko@gmail.com, $^{*}$corresponding author}
\end{center}

\vspace{2mm} \hspace{-7.7mm} \normalsize 
\keywordsformat{\textbf{Keywords:} heterophase particles, mechanical stress, magnetic states, 
 elongated nanoparticles, coatings, interfacial exchange interaction.}

\contentformat

\abstractformat{Abstract.} In this paper we investigate the dependence 
of the magnetic states of heterophase particles on mechanical multiaxial stresses. 
It is shown that for such particles, there are four possible states, 
and the conditions of stability of these states are determined.

\section{Introduction}

It can be assumed that the actual small magnetic particles are, for the most part, 
heterophase and not homogeneous. Such a conclusion can be drawn from the experimental fact: 
with decreasing of particle sizes, their reactivity increases, 
although the surface area of particles decreases.
The formation of neighboring magnetic phases can be caused by processes
of oxidation or disintegration of the solid solution 
(see, e.\,g.,~\cite{Stacey1974,Gapeev1992,Artemova1988}) occurring in the magnetically
ordered grain.
Sufficiently detailed theoretical study of the magnetic states and magnetization processes 
of the system of uniaxial particles presented in~\cite{Yang1991,Afremov1996,Afremov1999}.
In this paper we attempt to expand the model~\cite{Afremov1996} over multi-axis heterophase 
particles and to investigate the influence of mechanical stress on the magnetic state of such particles.

\section{Model}
\begin{figure}
\centering
\includegraphics[scale=0.8]{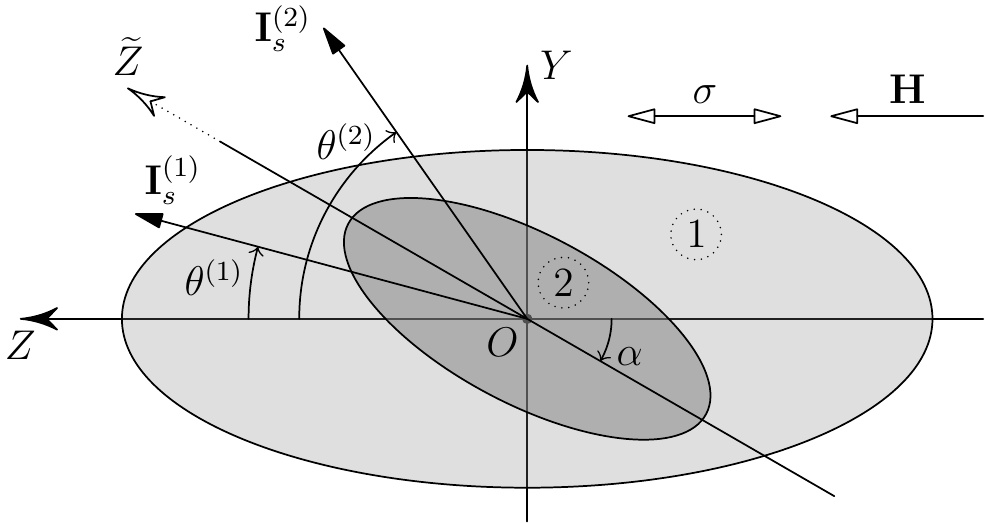}
\caption{Illustration of the model of two-phase particle}
\label{fig:1}
\end{figure}

\begin{enumerate}
\item  Homogeneously magnetized nanoparticle (phase {\em 1}) of volume $V$
has the form of ellipsoid with elongation\footnote{{\em elongation} – 
the ratio of the length $a$ of semi-major axis of the ellipsoid to the length $b$ of semi-minor one}  $q_1$, 
and its long axis oriented along the $Oz$-axis (see fig.~\ref{fig:1}).

\item Nanoparticle contains an uniformly magnetized ellipsoidal inclusion (phase {\em 2}) 
with a volume $v=\varepsilon V$ and elongation $q$.

\item It is considered that the axes of crystallographic anisotropy of both uniaxial ferromagnets are parallel 
to the long axes of the ellipsoids, and the vectors of spontaneous magnetization of 
phases ${{\mathbf I}}^{(1)}_s$ and ${{\mathbf I}}^{(2)}_s$ lie in the plane $yOz$, 
that contains the long axes of the magnetic phases, and make angles $\theta^{(1)}$ and $\theta^{(2)}$ 
with the $Oz$ axis, respectively.

\item Both external magnetic field $H$ and uniaxial mechanical stresses $\sigma $ are applied along the $Oz$-axis. 

\item The volume of nanoparticles exceeds the volume of superparamagnetic transition (that means that
we consider thermal fluctuations as negligible).
\end{enumerate}

We use the expression for the energy density of the nanoparticle that includes 
the contributions of the following types: 

\begin{enumerate}
\item energy density of anisotropy, which, according to~\cite{Chikazumi2009}, depending on the sign 
of the anisotropy constant has the form (all constants are defined below):
\be\label{eq1} 
\hspace{-2cm}
E_A=\left\{
  \begin{array}{l}
    \frac{1}{4}k^{(1)}_{A1}\left(I^{(1)}_s\right)^2
            \left(1-\varepsilon\right){\sin^{2}2\theta^{(1)}}+\frac{1}{4}k^{(2)}_{A1}\left(I^{(2)}_s\right)^2\varepsilon \sin^2 2\theta^{(2)}
    ,\ k^{(1)}_{A1}>0,\ k^{(2)}_{A1}>0, \\ 
\frac{1}{54}k^{(1)}_{A2}\left(I^{(1)}_s\right)^2\left(1-\varepsilon\right){\sin^{2} \theta^{(1)}}
    {\left(1+2{\cos{2\theta}^{(1)}}\right)}^2+\\ 
\quad+\frac{1}{54}k^{(2)}_{A2}\ \left(I^{(2)}_s\right)^2\varepsilon {\sin^{2} \theta^{(2)}}\left(1+2{\cos  2\theta^{(2)}}\right)^2,
\ k^{(1)}_{A1}<0,\ k^{(2)}_{A1}<0, \\ 
\frac{1}{4}k^{(1)}_{A1}\left(I^{(1)}_s\right)^2\left(1-\varepsilon \right){\sin^{2} 2\theta^{(1)}}+\\
    \quad+\frac{1}{54}k^{(2)}_{A2}\left(I^{(2)}_s\right)^2\varepsilon {{\sin }^{2} \theta^{(2)}}
        {\left(1+2{\cos  2\theta^{(2)}}\right)}^2,\quad k^{(1)}_{A1}>0,\ k^{(2)}_{A1}<0, \\ 
\frac{1}{54}k^{(1)}_{A2}\left(I^{(1)}_s\right)^2\left(1-\varepsilon \right){\sin^{2} \theta^{(1)}}{\left(1+2{\cos{2\theta}^{(1)}}\right)}^2+\\
    \quad+\frac{1}{4}k^{(2)}_{A1}\left(I^{(2)}_s\right)^2\ \varepsilon {\sin^{2} 2\theta^{(2)}},
    \ k^{(1)}_{A1}<0,\ k^{(2)}_{A1}>0,
        \end{array}
\right. 
\ee 

\item energy density of the magnetoelastic interaction with the field of uniaxial mechanical stress $\sigma$:
\be\label{2} 
E_{\sigma}=\frac{1}{2}{(I^{(1)}_s)}^2\left(1-\varepsilon \right){\Lambda}^{(1)}\sigma{\sin^{2} \theta^{(1)}}+
        \frac{1}{2}{(I^{(2)}_s)}^2\ \varepsilon {\Lambda }^{(2)}\sigma \sin^{2} \theta^{(2)}, 
\ee

\item energy density of the interaction of the magnetic moment with its own 
magnetic field, which can be represented as:
\be\label{eq3} 
\begin{aligned}
&E_m=
    -\frac{{{(I}^{(1)}_s)}^2}{4}
    \left(\left(1-2\varepsilon \right)k^{(1)}_N+\varepsilon k^{(2)}_N\right){\cos 2\theta^{(1)}}
    +\frac{{{(I}^{(2)}_s)}^2}{4}\varepsilon k^{(2)}_N{\cos 2\alpha}\,{\cos 2\theta^{(2)}}+\\
    &\qquad+\frac{\varepsilon}{3}{I}^{(1)}_s{I}^{(2)}_s\left.
        \left({-{\mathcal U}_1 \sin\theta^{(1)}{\sin\theta^{(2)}}}+{\mathcal U}_2{\cos\theta^{(1)}}{\cos\theta^{(2)}}
            \right)\right.,  
\end{aligned}
\ee

\item energy of exchange interaction across the border, which, according to~\cite{Yang1991}, can be defined as:
\be\label{eq4}
E_{ex} = -\frac{2A_{in}}{\delta}\cos\left(\theta^{(1)}-\theta^{(2)}\right)s.
\ee

\end{enumerate}

In \eqref{eq1} – \eqref{eq4} values $k^{(1,2)}_{Aj}=K_j/\left(I_s^{\left(1,2\right)}\right)^2,\ j=1,2$ 
– dimensionless constants of crystallographic anisotropy of the first or second order for each phase, 
$K_1$ and $K_2$ – constants of anisotropy of the cubic crystal of the first or second order,\\ 
$\left\{
\begin{array}{ll}
{\Lambda}^{(1,2)}=3\lambda_{100}^{(1,2)}/\left(I_s^{(1,2)}\right)^2\mbox{ for }k_{A1}^{(1,2)}>0,\\
{\Lambda}^{(1,2)}=3\lambda_{111}^{(1,2)}/\left(I_s^{(1,2)}\right)^2\mbox{ for }k_{A1}^{(1,2)}<0
\end{array}\right.$,
$\lambda_{100}^{(1,2)}$ and $\lambda_{111}^{(1,2)}$ – magnetostriction constants of phases,
$s$ – area of interphase boundary, 
$A_{in}$ – constant of interfacial exchange interaction,
$\delta $ – width of the transition region, that has the order of the lattice constant,
${\mathcal U}_j=j\left(k^{(2)}_N-k^{(1)}_N\right)+6\,s\,A_{in}/v\delta I_s^{(1)}I_s^{(2)},\ j=1,2$.

\section{Stable and metastable states}
We define the equilibrium state of a particle by minimizing the free energy density
$E = E_A+E_{\sigma}+E_m+E_{ex}$. 
Depending on the sign of the anisotropy constants, four systems of equations can be obtained.
The solutions of these systems define the equilibrium state of magnetic moment of the particle.

When $k^{(1)}_{A1}>0$ and $k^{(2)}_{A1}>0$:
\be\label{eq5}
\left\{
    \begin{array}{l}
        k_{A1}^{(1)}(1-\ve)\sin 2\theta^{(1)}\cos 2\theta^{(1)}+\\
        \qquad+
        \left[(1-2\ve)k_N^{(1)}+\ve k_N^{(2)}+\Lambda^{(1)}\sigma(1-\ve)\right]\sin\theta^{(1)}\cos\theta^{(1)}-\\
        \qquad-j\left[{\mathcal U}_1\cos\theta^{(1)}\sin\theta^{(2)}+{\mathcal U}_2\cos\theta^{(2)}\sin\theta^{(1)}\right]=0\\
        jk_{A1}^{(2)}\ve\sin 2\theta^{(2)}\cos 2\theta^{(2)}
            +j\ve\left(k_N^{(2)}+\Lambda^{(2)}\sigma\right)\sin\theta^{(2)}\cos\theta^{(2)}-\\
        \qquad-\left[{\mathcal U}_1\cos\theta^{(2)}\sin\theta^{(1)}+{\mathcal U}_2\cos\theta^{(1)}\sin\theta^{(2)}\right]=0\\
    \end{array}
\right.
\ee

When both $k^{(1)}_{A1}<0$ and $k^{(2)}_{A1}<0$:
\be\label{eq6}
\left\{
    \begin{array}{l}
        \frac{k_{A2}^{(1)}}{18}(1-\ve)\sin 2\theta^{(1)}\left(3-4\sin^2 2\theta^{(1)}\right)+\\
        \qquad+
        \left[(1-2\ve)k_N^{(1)}+\ve k_N^{(2)}+\Lambda^{(1)}\sigma(1-\ve)\right]\sin\theta^{(1)}\cos\theta^{(1)}-\\
        \qquad-j\left[{\mathcal U}_1\cos\theta^{(1)}\sin\theta^{(2)}+{\mathcal U}_2\cos\theta^{(2)}\sin\theta^{(1)}\right]=0\\
        j\frac{k_{A2}^{(1)}}{18}\ve\sin 2\theta^{(2)}\left(3-4\sin^2 2\theta^{(1)}\right)
            +j\ve\left(k_N^{(2)}+\Lambda^{(2)}\sigma\right)\sin\theta^{(2)}\cos\theta^{(2)}-\\
        \qquad-\left[{\mathcal U}_1\cos\theta^{(2)}\sin\theta^{(1)}+{\mathcal U}_2\cos\theta^{(1)}\sin\theta^{(2)}\right]=0\\
    \end{array}
\right.
\ee

When $k^{(1)}_{A1}<0$ and $k^{(2)}_{A1}>0$:
\be\label{eq7}
\left\{
    \begin{array}{l}
        \frac{k_{A2}^{(1)}}{18}(1-\ve)\sin 2\theta^{(1)}\left(3-4\sin^2 2\theta^{(1)}\right)+\\
        \qquad+
        \left[(1-2\ve)k_N^{(1)}+\ve k_N^{(2)}+\Lambda^{(1)}\sigma(1-\ve)\right]\sin\theta^{(1)}\cos\theta^{(1)}-\\
        \qquad-j\left[{\mathcal U}_1\cos\theta^{(1)}\sin\theta^{(2)}+{\mathcal U}_2\cos\theta^{(2)}\sin\theta^{(1)}\right]=0\\
        jk_{A1}^{(1)}\ve\sin 2\theta^{(2)}\cos 2\theta^{(2)}
            +j\ve\left(k_N^{(2)}+\Lambda^{(2)}\sigma\right)\sin\theta^{(2)}\cos\theta^{(2)}-\\
        \qquad-\left[{\mathcal U}_1\cos\theta^{(2)}\sin\theta^{(1)}+{\mathcal U}_2\cos\theta^{(1)}\sin\theta^{(2)}\right]=0\\
    \end{array}
\right.
\ee

Finally, when $k^{(1)}_{A1}>0$ and $k^{(2)}_{A1}<0$:
\be\label{eq8}
\left\{
    \begin{array}{l}
        k_{A1}^{(1)}(1-\ve)\sin 2\theta^{(1)}\cos 2\theta^{(1)}+\\
        \qquad+\left[(1-2\ve)k_N^{(1)}+\ve k_N^{(2)}+\Lambda^{(1)}\sigma(1-\ve)\right]\sin\theta^{(1)}\cos\theta^{(1)}-\\
        \qquad-j\left[{\mathcal U}_1\cos\theta^{(1)}\sin\theta^{(2)}+{\mathcal U}_2\cos\theta^{(2)}\sin\theta^{(1)}\right]=0\\
        j\frac{k_{A2}^{(1)}}{18}\ve\sin 2\theta^{(2)}\left(3-4\sin^2 2\theta^{(2)}\right)
            +j\ve\left(k_N^{(2)}+\Lambda^{(2)}\sigma\right)\sin\theta^{(2)}\cos\theta^{(2)}-\\
        \qquad-\left[{\mathcal U}_1\cos\theta^{(2)}\sin\theta^{(1)}+{\mathcal U}_2\cos\theta^{(1)}\sin\theta^{(2)}\right]=0\\
    \end{array}
\right.
\ee

\section{Results and discussion}

From the equations~\eqref{eq5} – \eqref{eq8} it follows that in the absence of magnetic field
parallel or antiparallel  orientation 
of the magnetic moments of the phases (when $\sin\theta^{(1)}=\sin\theta^{(2)}=0$)
correspond to the minimum of energy.
Maximum of energy corresponds to the orientation of the magnetic moments which satisfies the condition 
$\cos\theta^{(1)}=\cos\theta^{(2)}=0$.
The other solutions do not satisfy the condition of extremum. 
They can only be realized in the model with perpendicular (to the $Oz$-axis) phase distribution.

\newpage
Thus, as in the case of uniaxial anisotropy, in the absence of an external magnetic field 
heterophase particle can be in one of the following states:

\begin{center}
\begin{tabular}{|c|c|c|c|}
\hline
1&2&3&4\\
\hline
     {\includegraphics[scale=0.5]{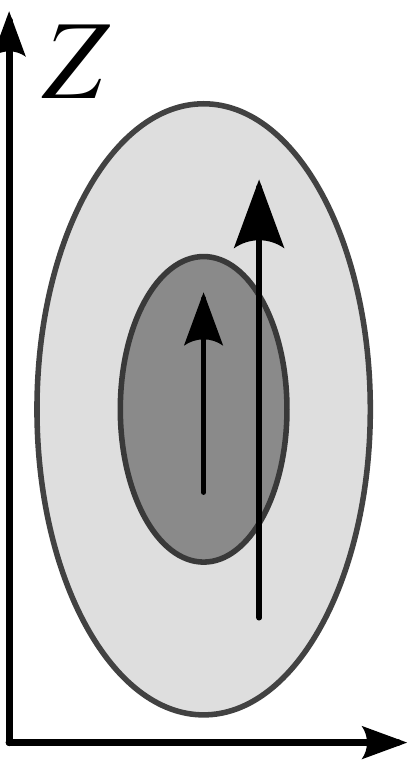}}
    &{\includegraphics[scale=0.5]{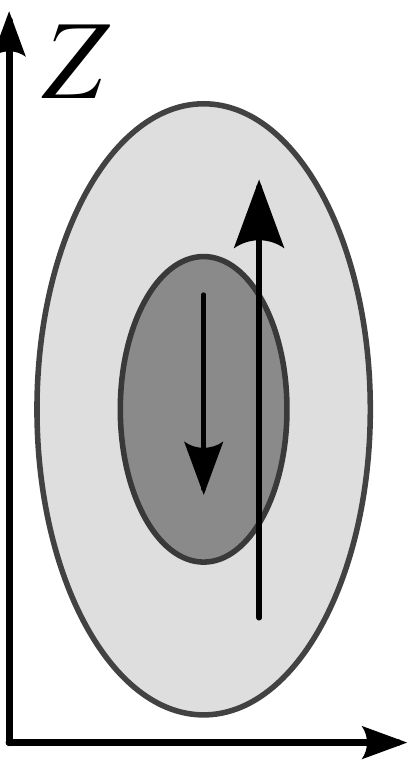}}
    &{\includegraphics[scale=0.5]{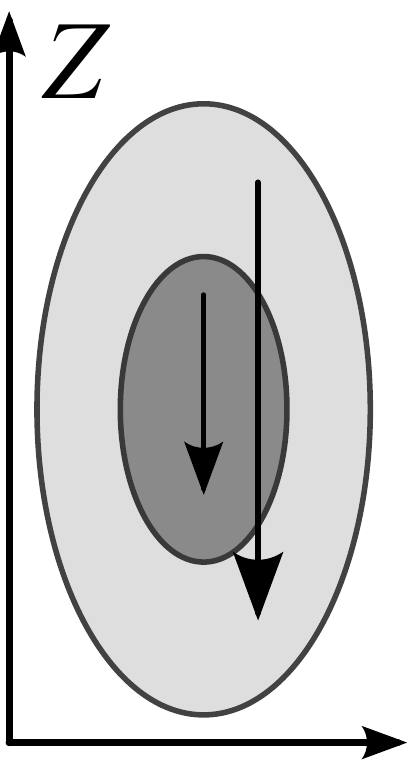}}
    &{\includegraphics[scale=0.5]{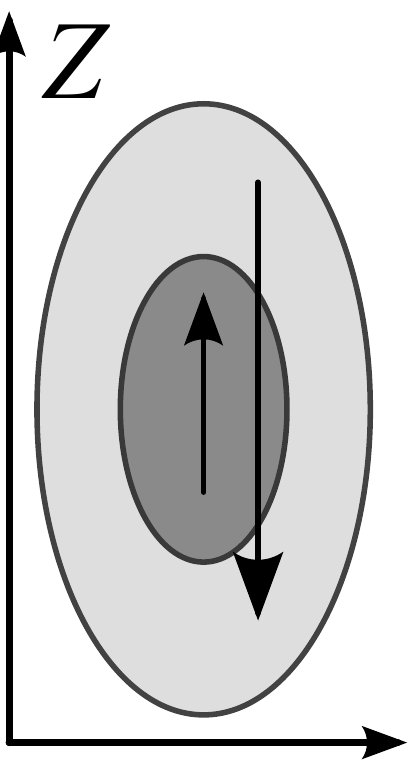}}\\%
\hline
\end{tabular}
\end{center}

If the magnetostatic interaction between the phases dominates the exchange interaction (${\mathcal U}_2>0$), 
the first and third states are metastable, 
since the free energy in these states than in the second and fourth.
Otherwise (${\mathcal U}_2<0$), the second and fourth are metastable state.

\section{Acknowledgments}
The work was supported by grants of Ministry of Education and Science: project 2.1.1/992 
«Structure and magnetic properties of nanocrystalline and amorphous multicomponent alloys»
and Federal Contract 02.740.11.0549, reference number 2010-1.1.-121-011,
«Magnetic properties and spin-transport phenomena in nanoscale condensed matter».

\bibliographystyle{ieeetr}
\bibliography{guangzhou}
\end{document}